# A Hyperspectral Microscope based on an Ultrastable Common-Path Interferometer

A. Candeo[1], B. E. Nogueira de Faria[2], M. Erreni[3], G. Valentini[1], A. Bassi[1],
A. M. De Paula[2], G. Cerullo[1], C. Manzoni[1]

[1] IFN-CNR, Dipartimento di Fisica, Politecnico di Milano, Piazza Leonardo da Vinci 32, I-20133 Milano, Italy
[2] Departamento de Física, Universidade Federal de Minas Gerais, 31270-901 Belo Horizonte-MG, Brazil
[3] Unit of Advanced Optical Microscopy, Humanitas Research Hospital, Via Rita Levi Montalcini 4, 20090, Pieve Emanuele, Milan, Italy

## Abstract

We introduce a wide field hyperspectral microscope using the Fourier-transform approach. The interferometer is based on the Translating-Wedge-Based Identical Pulses eNcoding System (TWINS) [Opt. Lett. 37, 3027 (2012)], a common-path birefringent interferometer which combines compactness, intrinsic interferometric delay precision, long-term stability and insensitivity to vibrations. We describe three different implementations of our system: two prototypes designed to test different optical schemes and an add-on for a commercial microscope. We show high-quality spectral microscopy of the fluorescence from stained cells and powders of inorganic pigments, demonstrating that the device is suited to biology and materials science. We demonstrate the acquisition of a 1Mpixel hyperspectral image in 75 seconds in the spectral range from 400 to 1100 nm. We also introduce an acquisition method which synthesizes a tunable spectral filter, providing band-passed images by the measurement of only two maps.

## Introduction

In the last decades, a huge advancement has occurred in optical microscopy, with revolutionary improvements in resolution and capability to reconstruct the three-dimensional features of samples, from nanoscale to millimetre size.[1,2] At the same time, a disruptive transformation has occurred in biological samples that optical microscopy is demanded to analyze. Genetically encoded probes, such as Green Fluorescent Protein (GFP), have been generated to target proteins, intra-cellular ions, metabolites, messengers ($H^+$, $Ca^{2+}$, $Cl^-$, etc.), in cells' and organisms' compartments. Moreover, molecular biologists have learned how to combine probes emitting at different wavelengths, either genetically expressed or based on immunohistochemistry, in the same specimen, in order to differentiate biological structures and reveal functional interactions. Remarkable examples of such probes are those operating on the basis of the Foster Resonant Energy Transfer (FRET), which exploits changes in the emission yield of the donor and the acceptor to measure the local concentration of cations ($Ca^{++}$), reactive oxygen species and other metabolites. As a matter of fact, these samples can be labelled in a full spectrum of colors and the combination of fluorescent probes emitting at several wavelengths is rather common. In addition, several fluorophores such as ratiometric dyes, voltage-sensitive dyes and pH sensitive dyes can be designed such that small variations of their absorption/emission spectra sensitively probe their local environment.

Similarly, in the field of material sciences, a number of novel materials present peculiar spectroscopic properties and resonance peaks in the visible and near infrared region. Among these, nanoparticles, quantum dots and two-dimensional semiconductor are constantly developed, exhibiting peculiar absorption and emission spectra. On one hand these optical properties are exploited for novel applications, on the other hand, the spectral information is key to understand the physical and chemical properties of the materials.

In this context, while a great work has been done in increasing optical resolution in microscopy, a still limited effort has been devoted to enhance spectral discrimination. In fact, when chromophores emitting in different spectral regions are measured, as in the case of FRET experiments, spectral discrimination is, generally speaking, still performed by a combination of bandpass filters and/or dichroic beam splitters. Yet, other options exist since the spectral information can be recovered with multispectral microscopes (MSMs) or hyperspectral microscopes (HSMs), which acquire a discrete (for MSMs) or continuous (for HSMs) absorption/emission spectrum for every pixel of the image, thus generating a so-called "spectral hypercube". Only few examples of MSMs and HSMs have been reported so far, the majority of which employs tunable filters, based either on liquid crystals or acousto-optic devices. This implementation can be obtained with a simple add-on to standard wide field microscopes; however, the acceptance angle and the transmission efficiency of wavelength selection elements are rather poor. MSMs are typically based on dispersive elements placed before the detection unit and often collect only a handful of spectral channels to reduce the acquisition time. The simplest approach to HSM is the so-called "whiskbroom", in which a confocal or multiphoton microscope reads out the spectrum while the excitation beam is scanned along the Field of View (FOV) in 2D.[3–7] The need to acquire a spectrum for every pixel severely limits the acquisition time. Hence, high performance whiskbroom configurations use a microlens array to massively parallelize the acquisition, in such a way to simultaneously record the spectra from thousands of points in the sample.[8] A more efficient approach relies on the pushbroom architecture, where the FOV is scanned line by line, while the spectra of all the pixels in the line are read out simultaneously.[9,10] This approach has been also implemented in a selective plane illumination microscopy to recover the spectral information from large 3D samples.[11] To go even faster, real-time snapshot microscopes have been proposed, which acquire the spectra of all the pixels in the FOV at once. Typically, this result is achieved through optical components that segment the FOV in many subregions, whose light is spectrally dispersed along the elements of a dense pixelated detector. Thus, the massive throughput of snapshot multispectral microscopes is obtained at a price of reduced spatial resolution[12–14] or high complexity.[15] Recent implementations of hyperspectral microscopes take advantage of the concept of compressive sensing, which has been originally inspired by the demonstration of the so-called single-pixel camera.[16] In practice, a digital micromirror device projects a sequence of patterns in the FOV of a wide field microscope while the light emitted or transmitted by the entire FOV is analyzed by a spectrometer. The spectral hypercube is then achieved through an appropriate inversion algorithm on the spectra measured from all the illumination patterns.[17]

The above-mentioned microscopes operate in the frequency domain and require a frequency selective device to decompose the radiation field into a number of spectral bands. An alternative HSM approach combines a monochrome imaging camera with a Fourier-transform (FT) spectrometer.[18] In FT spectroscopy an optical waveform is split by an interferometer in two delayed replicas, whose interference pattern is measured by a detector as a function of their delay. The FT of the resulting interferogram yields the intensity spectrum of the waveform.

FT spectrometers have remarkable advantages over dispersive ones: (i) higher signal-to-noise ratio in a readout-noise-dominated regime (the Fellgett multiplex advantage[19]); (ii) higher throughput, due to the absence of slits (the Jacquinot étendu advantage[20]); (iii) flexible spectral resolution, which is adjusted at will by varying the maximum scan delay.
When applied to an imaging system, the FT approach allows parallel recording of the spectra of all pixels within a two-dimensional FOV.

An FT-based imaging system must fulfil two key and challenging requirements: (i) the delay of the replicas must be controlled to within a fraction (1/100 or better) of the optical cycle (e.g., 0.02 fs at 600 nm); (ii) the bundle of rays that form the interferogram at a given pixel must have a high degree of coherence to add up constructively and to generate fringes with high visibility.

Few examples of HSMs based on the FT method have been proposed so far. Amongst the possible options, the Sagnac interferometer exploits the variable phase delay occurring between two counterpropagating beams that originate from the same radiation field.[21] Other implementations rely on the Michelson interferometer operating either on the illumination side[22] or on the detection side.[23,24] Based on this scheme, Müller et al. have implemented a Raman microscope using a proprietary optical device made of two corner cube reflectors operating in step-scan mode to control the Optical Path Difference (OPD) between the two arms. Waddudage et al. have proposed a very interesting implementation that relies on a nearly common path interferometer and produces an OPD uniform across the whole FOV. While remarkable features have been demonstrated and exciting applications have been proposed, the system remains rather cumbersome and difficult to operate in routine experiments.

Recently we introduced the Translating-Wedge-based Identical pulses eNcoding System (TWINS)[25,26], a common-path birefringent interferometer inspired by the Babinet–Soleil compensator, but conceived to provide retardation of hundreds of optical cycles between two orthogonal polarizations of a light beam. As the two fields share a common optical path, their phase delay can be adjusted with interferometric precision and displays an exceptionally high long-term stability, with phase jitter smaller than 1/360th of the optical cycle over 30 minutes, without active stabilization or tracking.[25] The TWINS interferometer has been successfully employed as a FT spectrometer in spectral ranges from the visible[27,28] to the mid-infrared[29,30], with both coherent and incoherent light beams. This approach combines the advantages of FT spectroscopy with the robustness and accuracy of a common-path interferometer. The TWINS has also enabled implementing a novel high-performance hyperspectral camera[31,32] and an imaging system for excitation and emission spectroscopy.[33]

In this paper we introduce a wide field HSM using the FT approach, in which the interferometer is based on the birefringent TWINS device. We describe three different implementations of our system: two prototypes designed to test different optical schemes and an add-on for a commercial microscope. We demonstrate that TWINS applied to FT microscopy provides a high degree of coherence at each pixel of the image, enabling HSM with short acquisition times and high spectral accuracy.

Design of the imaging system

A schematic of the TWINS interferometer is shown in Fig. 1(a). Blocks A and B are birefringent crystals made of α-barium borate (α-BBO), P1 and P2 are wire-grid polarizers with extinction ratio >5000 at wavelengths from 400 nm to 2 μm. P1 polarizes the input light at 45° with respect to the optical axes of A and B. Block A is a plate with thickness $L_A$ = 3.7 mm: it introduces, for each wavelength, a fixed phase delay between the two orthogonal polarizations that propagate along the fast and slow axes of the material. The optical axis of block B is orthogonal to the one of block A (see green double-arrow and circles in Fig. 1(a), respectively), thus introducing a delay of opposite sign. Hence, the overall phase delay introduced by the TWINS ranges from positive to negative values depending on the relative thicknesses of the two blocks. To enable fine tuning of the phase delay, block B is shaped in the form of two wedges with the same apex angle (7°), one of which is mounted on a motorized translation stage (L-402.10SD, Physik Instrumente) for lateral displacement with minimum motion step of 0.5 μm. Interference between the two replicas is guaranteed by polarizer P2, which projects the two fields to the same polarization at 45° with respect to the optical axes of A and B. In our TWINS design, the lateral translation of one wedge by 10 μm delays a wave at λ = 600 nm by 0.45 fs, equivalent to ≈1/4 of the optical cycle. As a consequence, an interferogram with total range of 1 ps, required to achieve a spectral resolution of 1THz (1.2nm at λ = 600 nm), is obtained by scanning the position of one wedge ±11 mm around the zero-path-delay (ZPD).

We will now discuss the suitability of the TWINS interferometer for FT imaging, and in particular for optical microscopy. In an FT imaging system, the interferometer is typically placed between the object and the detector plane. It is therefore basically different to employ TWINS as a spectrometer or for imaging: in the FT spectrometer[27,30] the light to be measured is collimated and propagates parallel to the main axis of the TWINS optical system; conversely, in an imaging system the rays from one object point O propagate at various angles through the interferometer before converging in the image spot. For this reason, it is necessary to evaluate the performance of the birefringent interferometer also taking into account the propagation direction of the light rays. Let's consider a generic ray that travels in the air at angle $\underline{\alpha} = (\alpha_x, \alpha_y)$ with respect to the interferometer optical axis, as defined in Fig. 1(c). Figure 1(b) shows the path of the ray through A and B; to simplify the visualization, the figure depicts the case in which $\alpha_y = 0$. Due to birefringence, the two orthogonally polarized replicas (red[blue]: normal[parallel] to the plane of the figure) travel along different optical paths and accumulate a total relative phase shift φ. The 2D map of Figure 1(d) shows φ estimated for incident directions $\underline{\alpha}$ in the range ±2.8°×±2.8°; the calculation was performed at the working conditions of our interferometer ($\lambda = 600$ nm, $L_A = 3.7$ mm) and in the case $L_B = L_A$, i.e. for a nominal delay of the interferometer for the ray travelling parallel to the main axis ($\underline{\alpha} = (0,0)$) set to 0. The lines overlapped to the 2D map are isophase levels, which show the hyperbolic pattern typical of birefringent imaging systems.[34]

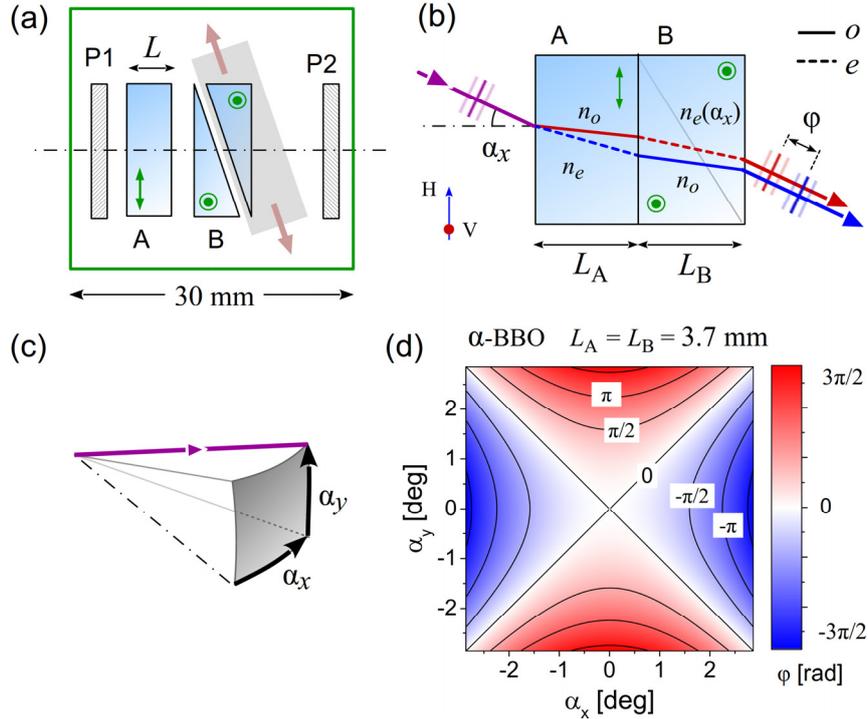

**Figure 1**: (a) Setup of the TWINS interferometer; P1, P2: polarizers; A-B birefringent blocks, made of α-BBO crystal; green double arrow and green circle: orientation of the optical axes. The dash-dotted line is the main axis of the TWINS optical system. (b) Optical paths of the vertically-polarized (red) and horizontally-polarized (blue) components of a ray traveling in the TWINS system; dashed lines: extraordinary ray; solid lines: ordinary ray. Spacing between the surfaces has been removed for simplicity; the drawing is a view from top, only showing angle $\alpha_x$. (c) Definition of angles ($\alpha_x$, $\alpha_y$), measured from the optical axis of the imaging system (dash-dotted line). (d) Relative phase φ accumulated by the crossed-polarized components as a function of ($\alpha_x$, $\alpha_y$).

Let's now consider an imaging system and follow the pencil of rays which propagate from one object point O to its image point on the detector plane, where they overlap giving rise to the interferogram. In the most general scenario, the rays of the bundle propagate through the interferometer at angles $\underline{\alpha}$ which span around the average value $\underline{\alpha}_0$ with range $\underline{\Delta\alpha}$, where $\underline{\alpha}_0$ and $\underline{\Delta\alpha}$ depend on the object point O, on the optical layout and on the position of the interferometer within the imaging system. As a consequence, the two replicas of each ray from O accumulate a phase shift φ($\underline{\alpha}$), whose values have a spread Δφ within the bundle. Such phase shift φ($\underline{\alpha}$) impacts on the resulting interferogram at the image point in two ways: (i) the visibility $v$ of its fringes degrades as the value of Δφ increases[31]; (ii) the ZPD shifts from the nominal zero of the interferometer to the average of the values of φ.

These general concepts provide the design criteria of a HSM based on the TWINS interferometer. In the following, we will introduce and discuss two wide-field HSM schemes: a "high-visibility" HSM, which maximizes the fringe contrast, and a "uniform-phase" HSM, which minimizes the ZPD excursion over the whole FOV. The second scheme has been also implemented by integrating the TWINS interferometer in a commercial microscope. For each scheme we characterize the interferometric properties and show HSM images of various test samples.

In all the HSM measurements discussed in the paper, each acquisition was performed by scanning the interferometer delay and acquiring one image for each delay. The maximum delay of the scan was adjusted to achieve the desired spectral resolution, while the sampling step was chosen according to the application, taking into account that oversampling with respect to the Nyquist limit improves the S/N ratio of the acquisition, but also increases the total acquisition time.

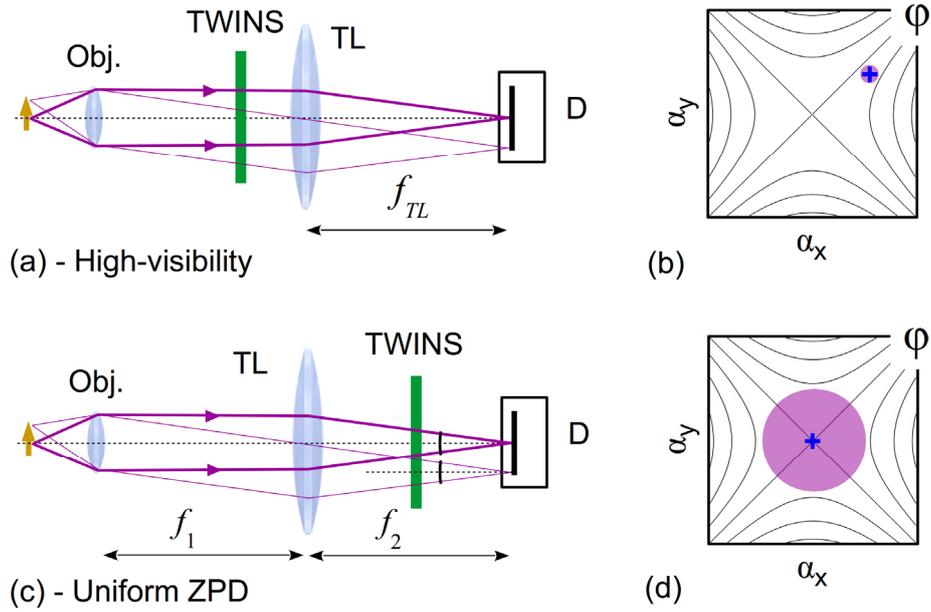

**Figure 2**: Optical configurations of the HSM discussed in the text. (a) High-visibility HSM. (b) Corresponding schematic map of φ (magenta area) accumulated by the bundle of rays from one point of the object. (c) Uniform-ZPD HSM. (c) Corresponding schematic map of φ (magenta area) accumulated by the bundle of rays from one point of the object. Obj.: objective; TL: tube lens; D: 2D detector; green line: TWINS interferometer.

*High-visibility HSM*

The first HSM scheme is illustrated in Fig.2(a): it consists of an infinite-corrected objective, a tube lens with focal length $f_{TL}$ = 200 mm and the 2D detector D. The birefringent interferometer is located between the microscope objective and the tube lens. In this configuration, all the rays from one object point are *parallel*, hence they all propagate at the same angle ($\Delta\pmb{\alpha}$ = 0; $\pmb{\alpha}$ = $\pmb{\alpha_0}$). This is illustrated in the schematic phase map of Fig. 2(b): since all rays of one object point travel at the same angle $\pmb{\alpha_0}$ (neglecting diffraction) they are all located in one point of the phase map (magenta area). Hence, the coordinate $\pmb{\alpha_0}$ of the average direction (blue cross) depends on the position of O within the FOV. In other words: (i) all the rays from the same object point experience a common phase shift φ (Δφ = 0) hence the visibility of the interferogram is maximum. For this reason, this configuration is labelled *high-visibility HSM*; (ii) the ZPD at each point of the image is φ($\pmb{\alpha_0}$), whose value varies throughout the FOV.

The detector is a monochrome EM CCD silicon camera (Luca R, Andor, Belfast, Northern Ireland) with size 8x8 mm$^2$, 1002×1004 pixels, 14-bits depth and spectral sensitivity from 400 nm to 1100 nm. In order to characterize the interferograms, we measured the surface of a white Spectralon (Labsphere Inc., North Sutton, USA), in reflection configuration. The light source was a broadband white LED, and the collection was performed by a 20x objective (Nikon, NA = 0.45). We acquired 600 images with wedge translation-steps of 2 μm. The main panel of Figure 3(a) shows the interferogram of the image pixel at the center of the FOV (pixel A of Fig.3(b)).

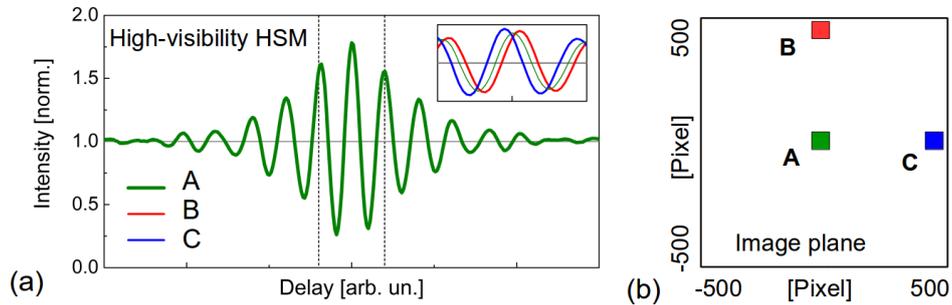

**Figure 3**: (a) Raw interferogram of the pixel at the center of the 2D camera, for the high-visibility HSM configuration; inset: expanded interferograms of three pixels (A, B, C) of the 2D detector, in the delay interval delimited by the vertical dashed lines of the main panel. (b) Detector plane; A, B, C are the pixels corresponding to the interferograms of panel (a).

An expanded view at early delays around the ZPD is shown in the inset, together with the interferograms at pixels B and C shown in Fig. 3(b). As expected, the three traces have different ZPD. By analyzing the experimental interferograms at each image pixel we calculated the ZPD and the visibility throughout the image plane. The ZPD at each camera pixel is shown in Figure 4(a), in terms of phase shifts for a wave at 600 nm; a section is shown in Figure 4(d).

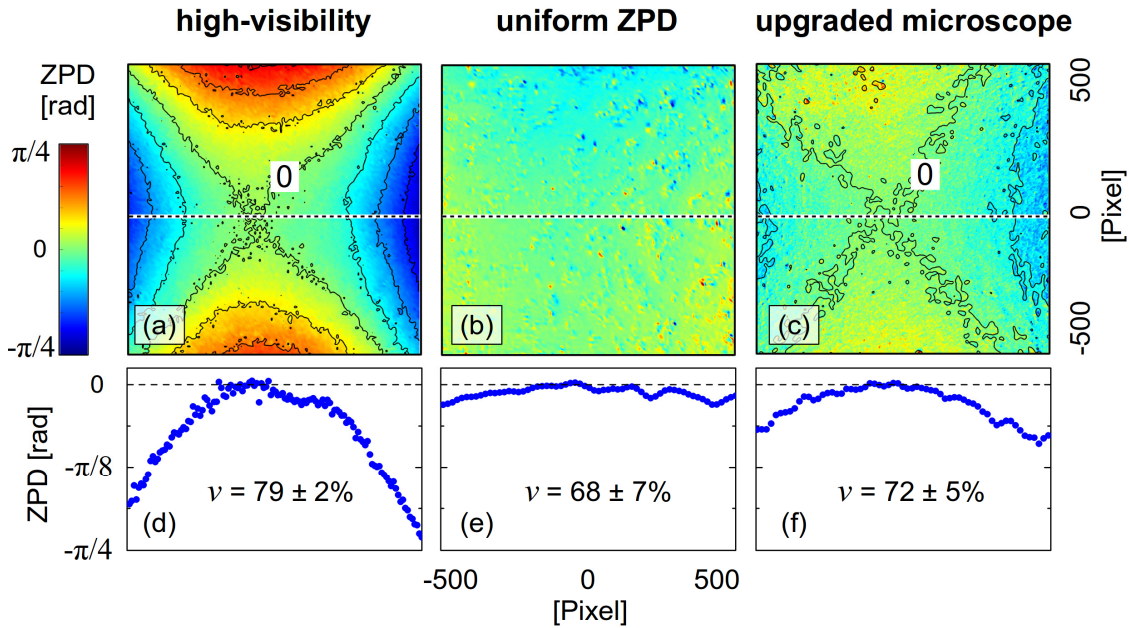

**Figure 4**: Measured ZPD for the three configurations of the HSM. (a-c) Maps of the ZPD at each camera pixel, displayed in the same color scale and in units of the phase shift of a wave at 600 nm wavelength; (a) high-visibility HSM, (b) uniform-phase HSM, (c) commercial microscope upgraded to FT hyperspectral imaging. The solid lines of panels (a) and (c) are isodelay data-points, at intervals of π/16. (d-f) ZPD at horizontal sections of the corresponding maps, at the coordinates marked by the dotted lines; the numbers are the visibility over the whole FOV.

The black lines of panel (a) are isodelay levels at intervals of π/16; their hyperbolic pattern recalls the isophase lines in figure 1(d). The ZPD ranges from -π/4 to +π/4, the largest phase shift being at the border of the FOV, i.e. in points B and C of Fig. 3(b). In this scheme, the ZPD excursion only depends on the maximum values of $\boldsymbol{\alpha}_0$, determined by the focal length $f_{TL}$ of the tube-lens and the size of the detector, and independent from the objective magnification. The average fringe visibility throughout the whole FOV is 79 +/- 2%; this value is lower than the expected 100% due to the non-optimal extinction ratio of the polarizers. However, it is worth noting that the reduction in contrast does not affect the quality of the hyperspectral images, except for a slight decrease in Signal to Noise Ratio (S/N). This effect can be compensated for by using a low-noise, high-dynamic-range pixelated detector, either CCD or CMOS, which is rather common in scientific applications.

By comparing the experimental delays of Fig. 4(a,d) with the phase shift of Fig. 1(d), we estimate that $\pmb{\alpha_0}$ range is ±1.12°×±1.12°, in close agreement with the value of ±1.14°×±1.14° expected from the detector size (8x8 mm$^2$) and tube-lens focal length ($f_{TL}$ = 200 mm) of our system.

As a first test of the high-visibility HSM, we imaged a high resolution RGB LCD screen. All emitters of the screen were activated with the highest power; the image was collected with a 10x objective. We acquired 600 images with sampling step of 2 μm; the total wedge excursion was hence +/-600 μm, leading to spectral resolution of 18 THz, sufficient to capture all the spectral features of the sample; the single-frame exposure was 140 ms, and the acquisition time of the image set, including the non-optimized lead-time for the electronic communications, was 240 seconds. Note that here we opted for over-sampling; the acquisition time reduces by a factor of 10 by sampling the same range at the Nyquist-Shannon limit (i.e. with a sampling step of ≈20 μm). The spectral results are summarized in Fig. 5; Figure 5(a) shows the color image reconstructed from the spectral hypercube after FT, while panels (b-d) display the spectra of each emitter. The solid lines are obtained from the HSM, while the dashed curves are measured with a conventional grating spectrometer, after setting the screen to the primary colors one at a time.

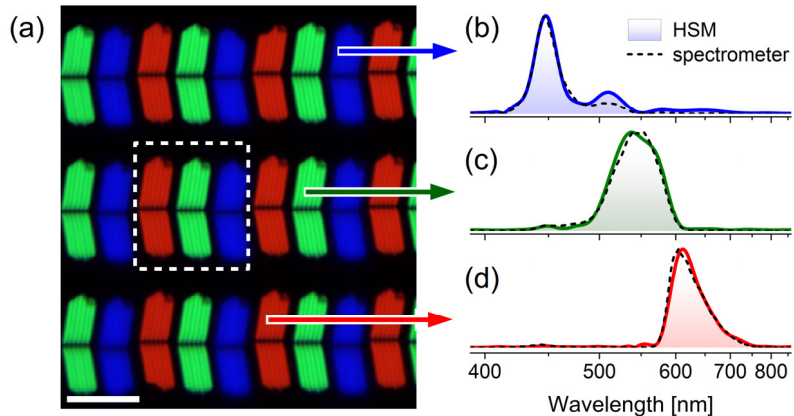

**Figure 5**: HSM image of an RGB LCD screen, acquired with the high-visibility HSM. (a) RGB image, synthesized from the spectral hypercube. Scale bar is 50 μm. Dashed line: screen pixel. (b-d) Normalized spectra of each individual emitter. Solid line: spectrum retrieved with the high-visibility HSM; dashed line: average spectrum collected by a standard grating spectrometer, when only red, blue and green emitters of the screen are activated.

*Uniform-ZPD HSM*

The second HSM scheme is illustrated in Fig.2(c): in this case, the birefringent interferometer is placed between the tube lens and the detector D. The rays of the bundle from one object point reach the TWINS interferometer with variable angles $\pmb{\alpha}$; in the schematic phase map of Fig. 2(d), the range of values of $\pmb{\alpha}$ is represented by the magenta area, the average angle $\pmb{\alpha_0}$ is the blue cross. If $f_1 = f_2$, the average value is $\pmb{\alpha_0} = \pmb{0}$ for all object points. In this configuration: (i) the ZPD at each point of the image is φ(**0**)=0, and is hence space invariant; for this reason, this configuration is labelled *uniform-ZPD HSM*; (ii) all the rays from one object point accumulate a non-uniform phase shift (Δφ ≠ 0), which reduces the visibility of the interferogram at all the image points.

In order to characterize this scheme, we employed the same objective, tube lens and camera used with the high-visibility HSM, making sure that $f_1 = f_2 = f_{TL}$ to obtain $\pmb{\alpha_0} = \pmb{0}$. From the measurement of the white Spectralon, we verified that the visibility reduces to 68%, while the ZPD is uniform throughout the FOV, as shown in Figs. 4(b,e). Despite the reduced interferometric contrast, the uniform ZPD offers a unique advantage in FT imaging, related to the sampling strategy. Typically, an unknown signal is acquired at a uniform sampling rate, whose step size must fulfil the Shannon/Nyquist sampling theorem to avoid spectral aliasing. However, in some cases other sampling strategies can be followed, such as spectral compressive sensing[35] when the spectrum to be measured can be reduced to a sparse representation. Most of these approaches require that data are acquired at specific sampling positions. Due to the central role of sampling delay, these techniques can be extended to FT imaging only if the delay imposed by the interferometer is independent of the spatial position of the pixels. Hence, our *uniform-ZPD HSM* scheme enables the application of such sampling strategies. Taking advantage of this peculiar property, we also developed a novel narrowband detection approach, which will be discussed in the following sections.

*Upgraded commercial microscope*

The uniform-ZPD optical scheme was used to upgrade to HSM a commercial optical microscope (Leica DMRBE). The TWINS interferometer was placed between the tube lens and the 2D detector of the microscope. In this case, the visibility of the interferogram resulting from HSM acquisition of the white Spectralon is 72%. The ZPD is shown in Figure 4(c,f); it ranges from $-\pi/16$ to $+\pi/16$, which we ascribe to the imperfect match of $f_1$ with $f_2$ due to mechanical constraints posed by the microscope chassis.

We tested the upgraded commercial microscope by imaging three samples: an RGB AMOLED display, doubly-labelled cells, and pigment powders.

Figure 6 shows the image of the display, when all emitters were activated. The image was collected with a 20x objective; exposure time was 100 ms per frame, and we acquired 240 frames (5μm wedge step, ±600 μm scan). The total measurement time, including the electronic lead time was 75 seconds. The RGB image of panel (a) is a reconstruction from the spectral hypercube; panels (b-d) show the average spectra of each emitter extracted from the hypercube (solid lines), together with the spectra measured with the conventional grating spectrometer (dashed lines) from the whole display.

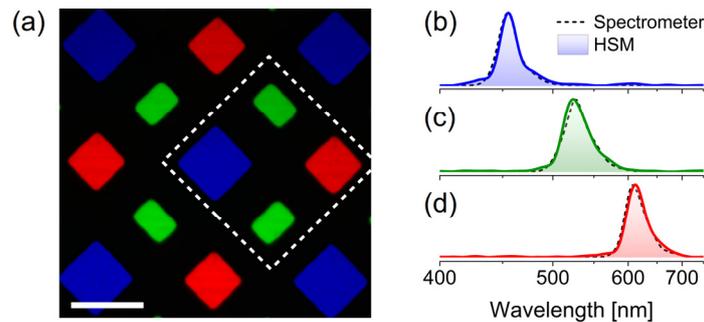

**Figure 6**: HSM image of an RGB AMOLED screen, acquired with the upgraded microscope. (a) RGB image, synthesized from the spectral hypercube. Scale bar is 50 μm. Dashed line: screen pixel. (b-d) Normalized spectra of each individual emitter. Solid lines: spectra retrieved with the high-visibility HSM; dashed line: average spectra collected by a standard grating spectrometer, when only red, blue and green emitters are activated. The axis is uniform in frequency.

The high spectral accuracy of our scheme enables a precise spectral analysis of the image. Careful inspection of Fig. 6(a) shows that the emission of each color is not uniform across the sample. By automatic *k*-means image segmentation[36], the surface of each emitter was separated into 3 classes with distinct spectra; the pseudocolor image of Fig. 7(a) shows the clusters, clearly revealing the presence of a substructure, barely visible in the RGB image of Fig. 6(a). The normalized spectra of the three classes exhibit variations of the order of only few percent; to visualize them, in Fig. 7(b) we present the differences between the normalized spectra within each emitter. The horizontal line indicates the reference spectrum and its color indicates the cluster that was taken as reference.

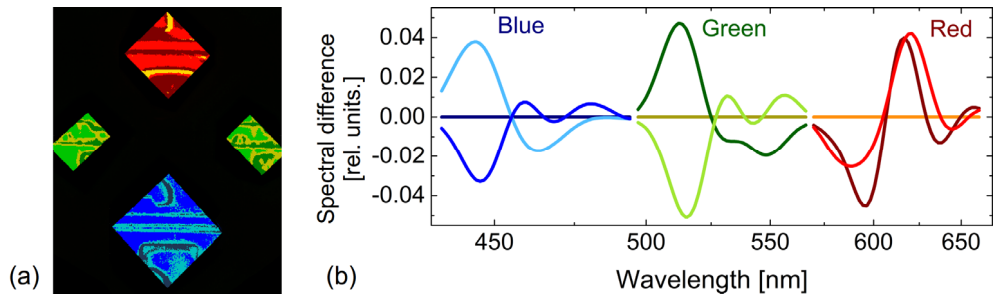

**Figure 7**: *k*-means segmentation of the RGB AMOLED screen. (a) pseudocolor image of one pixel unit; areas with the same color have emission spectra with the same shape; (b) difference between the normalized spectra of the clusters, plot with the same color codes used in panel (a). The horizontal line indicates the spectrum and color of the cluster that is taken as reference. The axis is uniform in frequency.

Figure 8 shows an HSM image of cells from the human colorectal adenocarcinoma cell line Caco2. Membranes are stained by Alexa Fluor[TM] 430, nuclei by Alexa Fluor[TM] 488. The images were acquired in epifluorescence using a Hg lamp (Osram HBO 50W) and a Leica I3 filter cube (excitation: bandpass at 450-490 nm; dichroic mirror edge: 510 nm; emission: long pass at 515 nm). Fluorescence images were taken with a 60x objective (Olympus UPLFLN NA=0.9). We collected 150 frames with 400-ms exposure time each, and with 8-μm wedge translation steps (±600 μm scan). With these parameters, the sample exhibited negligible photobleaching. Figure 8(a) shows the pseudocolor image deduced from the hypercube. Panel (b) displays the fluorescence spectra of nuclei and the membranes, after the subtraction from the whole hypercube of the common background (shaded area), obtained averaging the spectra of the dark areas of the image. The measured spectra are in very good agreement with the fluorescence spectra expected from Alexa Fluor[TM] 430 and 488 when excited slightly off-band as in our case; note that, although the emission from the membrane is significantly smaller (c.a 10 times weaker) than the signal from the nuclei, the S/N ratio of the HSM enables discriminating the two contributions.

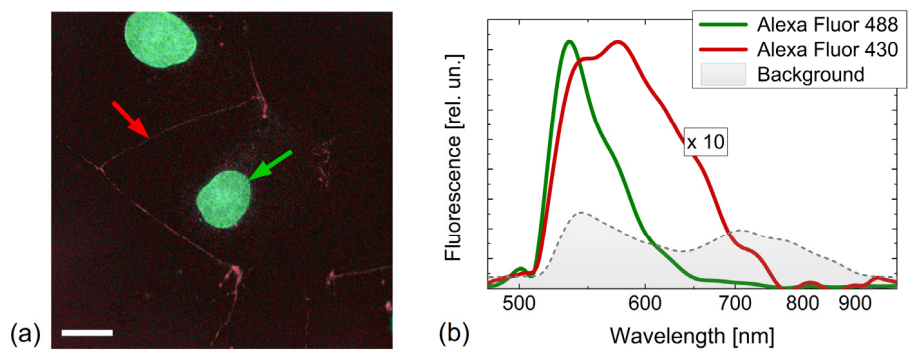

**Figure 8**: Hyperspectral image of colorectal adenocarcinoma cell line Caco2, acquired with the upgraded microscope; membranes are stained by Alexa Fluor[TM] 430, nuclei by Alexa Fluor[TM] 488. Excitation: Lamp. (a) false-colour image synthesised from the spectral hypercube; scale bar is 20 μm. (b) Shaded area: spectrum of the common background; solid lines: Fluorescence spectra from the two points indicated by the arrows in panel (a), after the subtraction of the common background. Fluorescence from Alexa Fluor[TM] 430 has been expanded 10x for clarity. The axis is uniform in frequency.

Beyond molecular biology, materials science is an important field of applications of HSM. In fact, the response to light, in terms of absorption and emission spectra, is a practical and effective way to investigate the properties of materials, including semiconductors and low-dimensional materials, where quantum confinement determines the electronic and optical properties. To test our microscope in this scenario we prepared a mixture of powders made of two inorganics pigments: Egyptian Blue and Cadmium Orange, which owe their colors to photophysical properties. Egyptian Blue is a synthetic pigment made of Cuprorivaite ($CaCuSi_4O_{10}$), first produced in the 4th Dynasty in Egypt (around 2500 BC) and widely used in the Mediterranean basin until the end of the Roman period and beyond.[37] When excited with red light, it exhibits an exceptionally strong infrared emission, such that nowadays it has been considered for cooling of sunlit surfaces. Cadmium Orange belongs to a class of pigments based on Cadmium Sulphide (CdS) and Cadmium Sulphoselenide (Cd(S,Se)) that were introduced in the second half of the 19th century and largely used since then, until concerns about their toxicity have been raised. These pigments have typical semiconductor characteristics and owe their hue to the energy gap. Beyond being pigments, Cadmium Sulphide and Cadmium Selenide have been largely employed in photodetectors. Time-resolved photoluminescence has shown that, when excited with light of wavelength below the cutoff, they exhibit a faint and very fast visible emission. This indicates that free carriers quickly relax to trap states, from which they recombine emitting a rather intense long living fluorescence in the NIR range.[38] For the experiments finely ground pigments purchased from Kremer Pigmente GmbH ($CdS_{1-x}Se_x$ x = 0.17 - Cadmium Orange No. 0.5, Egyptian Blue) were mixed and dispersed on a microscope slide. Excitation was obtained with the Hg lamp using the Leica I3 filter cube. We acquired 300 frames with 400-ms exposure time each, and with 4-μm wedge translation steps (±600 μm scan). Figure 9(a) shows the false-color image synthesized from the hyperspectral cube, while panel (b) depicts the fluorescence spectra of CdSSe and Cuprorivaite measured with the HSM in the positions indicated by arrows in panel (a). As expected, the fluorescence only consists of the NIR component, which is ascribable to a symmetrically prohibited transition of $Cu^{2+}$ in the case of Cuprorivaite and to trap state relaxation in the case of Cadmium Sulphoselenide.

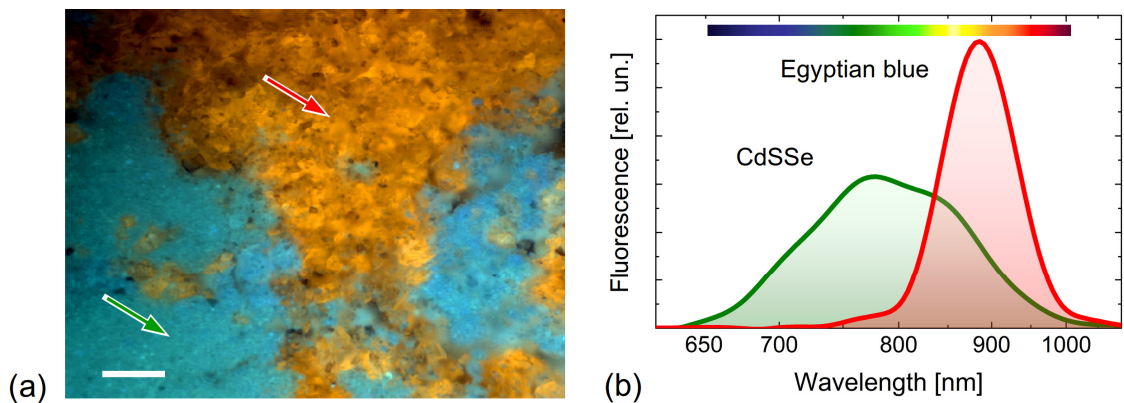

**Figure 9**: Hyperspectral fluorescence false-color image of a mixture of $CdS_{1-x}Se_x$ (x = 0.17) and Egyptian blue powders, acquired with the upgraded microscope. (a) False-color image, synthesized from the hyperspectral cube; scale bar is 100 μm. (b) Fluorescence spectra of CdSSe and Egyptian blue measured with the HSM microscope at the areas indicated by the arrows. Colorbar: false-color spectral assignment.

**Synthetic tunable filter**

The FT approach enables the retrieval of the broad spectrum of a radiation field by a time-domain scan: in practice, it is the reciprocal of the frequency-domain acquisition by a sequence of bandpass filters. For direct acquisition of images at a discrete number of spectral bands, the latter technique is better suited than the scanning FT method, since it requires the collection of fewer frames, one for each band. As illustrated in Figure 4, the *uniform-ZPD HSM*, in both the prototype form (panel (b)) and the upgraded microscope (panel (c)), offers the unique advantage that the delay imposed by the interferometer has very little variation within the image plane. This enables the application of those techniques (such as compressive sensing strategies) which require that all data are acquired at specific sampling positions, independent from image coordinate.

In this section, we introduce an acquisition method which is based on the FT approach and on the accurate choice of the sampling positions; when applied to the uniform-ZPD HSM, it synthesizes a *tunable spectral filter*, providing images around a given frequency $f_0$ by the acquisition of only two maps. The two maps are named In-Phase and Quadrature; each of them is obtained by acquiring images at well-defined sampling positions (e.g. by modulating the illumination or switching On-Off the detection optical path) and integrating them directly in the detector by the multiple exposure method.[39] The sampling is with period $T = 1/f_0$ (i.e. the optical cycle of the component at frequency $f_0$) and is referenced to the ZPD of the interferogram. The In-phase map corresponds to sampling at even multiples of T/2, while Quadrature collects at odd multiples of T/2. The band-passed image at $f_0$ is simply obtained by taking the difference between In-Phase and Quadrature maps. The synthetic filter can be understood in the time domain: at the In-Phase delays (even multiples of T/2, i.e. multiples of T), *constructive* interference occurs for the waves at frequency $f_0$; conversely, at the Quadrature delay (odd multiples of T/2), *destructive* interference occurs at frequency $f_0$. The information at $f_0$ is completely absent in the Quadrature image; on the other hand, the contribution of the information at all the wavelengths other than $f_0$ is present both in the In-Phase and Quadrature images. The subtraction of the Quadrature from the In-Phase image provides information at $f_0$ and rejects all the wavelengths other than $f_0$. A more rigorous derivation of the method in the frequency domain, together with the effective band of the filter, is provided in Appendix 2.

This new protocol has two main advantages: (i) it needs only two maps (In-Phase and Quadrature) instead of the numerous (≈ 100) images required by the hyperspectral detection; (ii) the multiple exposure mechanism allows one either to lower the illumination dose, or to reduce the acquisition time at each delay to tens of milliseconds, still exploiting the full camera dynamic range. In fact, each of the two maps is downloaded from the camera only once, when the dynamic range of the detector has been filled up, thus making negligible the readout noise, which is typically the major noise source in scientific grade cameras.

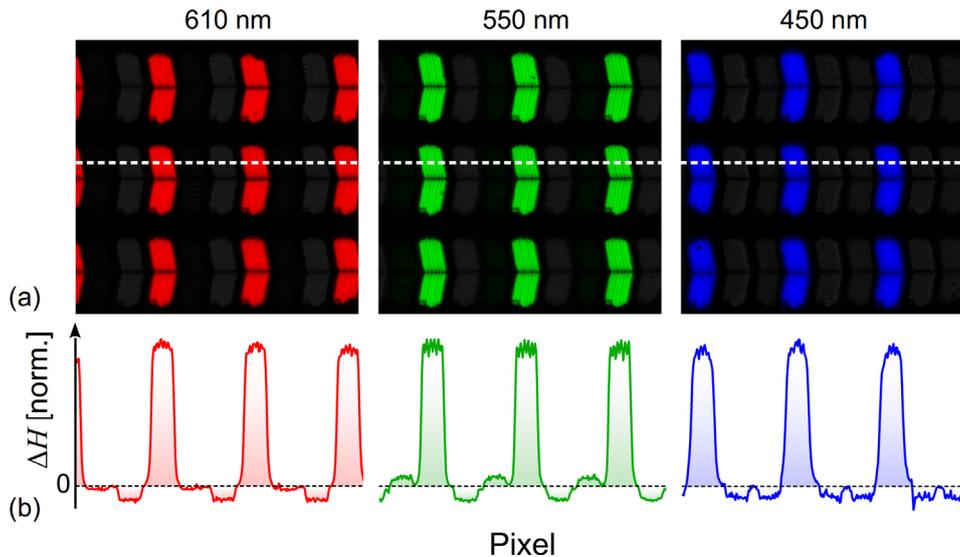

**Figure 10**: band-passed images of the RGB LCD screen, acquired with the upgraded microscope and following the synthetic-filter approach for N = 20. (a) False-colours images of the screen, filtered at 450, 550 and 610 nm. (b) Spatial intensity of image sections corresponding to the white dashed lines. We used this method with our upgraded microscope to collect 3 band-passed images of the RGB LCD screen discussed in Figure 5. As in that case, the images were acquired when all R, G and B

emitters where simultaneously radiating; the filters were set at wavelengths corresponding to the spectral peak of each emitter, namely 450, 550 and 610 nm. Each In-Phase and Quadrature image was the multiple exposure acquisition of N = 20 samples. The resulting band-passed images are depicted in Fig. 10(a), and separately show the isolated blue, green and red emitters. Figure 10(b) shows the normalized spatial intensity of image sections corresponding to the white dashed lines. The negative values of the intensity are due to the shape of the spectral filter, which has also negative tails (see Fig. 11(d) in Appendix 2).

## Conclusions

We have demonstrated that the TWINS interferometer is well suited to design a very compact hyperspectral microscope featuring: (i) wide spectral coverage; (ii) user adjustable spectral resolution; (iii) spatial resolution basically unaffected compared to a standard microscope; (iv) compactness; (v) high spectral sensitivity thanks to the large contrast of the interferometer.

The exceptional compactness of the TWINS interferometer allowed us to develop two implementations of the hyperspectral microscope, which have been thoroughly characterized: the high-visibility HSM, characterized by large interferometric contrast, and the uniform-ZPD HSM, providing space-invariant detection. Both prototypes demonstrate the capability of acquiring image sets with a large number of spectral bands. In the two prototypes, each one of the main properties (i.e. interferometric contrast and space invariance) have been pushed to maximum, in order to tailor the microscope for different applications. When space invariance is not a concern, maximum contrast, to fully exploit the dynamic range of the camera, is achieved by placing the TWINS between the objective and the tube lens. Conversely, when compressive sensing approaches or smart sampling schemes have to be applied, space invariance requires the TWINS to be placed between the tube lens and the camera. The uniform-ZPD system was also applied as an add-on module to upgrade a commercial microscope. The device has been employed to address two typical problems in biology and materials science and has demonstrated the capability of measuring the spectral features of the analysed samples producing highly informative datasets. We demonstrated the acquisition of 1Mpixel hyperspectral images with spectral resolution of 18 THz (22nm at $\lambda = 600$ nm) in less than 75 seconds. By increasing the acquisition time, the resolution can be pushed to 1 THz, only limited by the interferometer scan range. For the uniform-ZPD HSM, we also introduced an acquisition method to synthesize a tunable spectral filter, providing band-passed images by the measurement of only two maps.

## Appendix 1 – Human cells preparation

The human colorectal adenocarcinoma cell line Caco2 was cultured on glass slides coated with Poly-L-Lysine, in RPMI 1640 medium supplemented with 10% fetal bovine serum, 2mM Ultraglutamine-1 and 100U/ml penicillin/streptavidin. Cells were then fixed in 4% PFA, 15 min at RT. After two washes in Phosphate Buffered Saline with Calcium and Magnesium (PBS$^{+/+}$), cells were incubated overnight at 4°C with the specific primary antibodies against Lamin A/C (SantaCruz 7292-X, for the nuclei) and Zona occludens-1 (ZO-1, Invitrogen 61-7300, for the membrane), diluted in 2% bovine albumin serum (BSA), 0.1% TritonX-100, 5% normal goat serum in PBS$^{+/+}$. After 3 washes in washing buffer (0.05% Tween20 in PBS$^{+/+}$), cells were incubated with Alexa Fluor 430- and Alexa Fluor 488-conjugated, species-specific cross-adsorbed detection antibodies, in washing buffer for 1h at RT, in order to visualize ZO-1 and Lamin A/C, respectively. After 4 washes in washing buffer, slides were mounted with Mowiol solution.

# Appendix 2 – The synthetic tunable filter

The synthetic tunable filter is schematically illustrated in Fig. 11; it refers to the acquisition at only one image point, since extension to imaging is straightforward in our HSM, as will be explained later. $g(t)$ (panel (a)) is the interferogram at the point, and $G(f)$ (panel (b)) is its FT; this is a real and even function, and its side lobes are the intensity spectra of the field originating the interferogram; $s(t)$ (panel (c)) is the sampling function, which has uniform sampling step $T$ and is shifted by $t_0$ with respect to the ZPD of the interferogram. If we now define $f_0 = 1/T$ and assume that no spectral aliasing is occurring (which can be obtained by optically filtering high spectral frequencies, equivalent to higher orders in a grating spectrometer) the multiple exposure corresponds to *integrating* the sampled function $g(t)s(t)$. Sampling $g(t)$ at rate T creates, in the frequency domain, a periodic summation of $G(f)$ shifted by multiples of $f_0$. In addition, the value of the integral in time of a function is the amplitude of its FT at $f = 0$. Hence, after sampling and integrating, one obtains:

$$H = \int g(t)s(t)\, dt = G(0) + 2G(f_0)\cos(2\pi f_0 t_0), \qquad (1)$$

where the cosine term accounts for the time offset $t_0$. Eq. (1) shows that the value H includes $G(0)$, which is proportional to the average value of the interferogram, and a term proportional to the spectral component $G(f_0)$. In particular, by choosing $t_0 = 0$ and $t_0 = T/2$, we obtain two values of H, which we term In-Phase and Quadrature respectively:

$$\text{In-Phase } (t_0 = 0): H_P = G(0) + 2G(f_0) \qquad (2a)$$

$$\text{Quadrature } (t_0 = T/2): H_Q = G(0) - 2G(f_0) \qquad (2b)$$

Note that the In-Phase case corresponds to sampling at even multiples of T/2, while the Quadrature collects at odd multiples of T/2. By taking the difference $\Delta H$ between the In-Phase and Quadrature values, we get

$$\Delta H = H_P - H_G = 4G(f_0), \qquad (3)$$

thus subtracting the $G(0)$ contribution. The method implements a synthetic spectral filter at $f_0$ by simply storing two datapoints, namely $H_P$ and $H_G$. From Eq.(1) we can also estimate the robustness of the method when the delay $t_0$ is erroneously shifted to $0+\Delta t$ and $T/2+\Delta t$ during the detection of the In-Phase and Quadrature datapoints, respectively. In this case, the difference $\Delta \widetilde{H}$ would be:

$$\Delta \widetilde{H} = 4G(f_0) \cdot \cos(2\pi f_0 \Delta t) = \Delta H \cdot \cos(2\pi f_0 \Delta t), \qquad (4)$$

corresponding to a reduction of the detected signal. According to Eq. (4), an erroneous delay $|\Delta t| = T/8$ (corresponding to phase delay of $\pm\pi/4$) would reduce the signal to $\Delta H/\sqrt{2}$. Note that $\pm\pi/4$ is the maximum shift of the interferogram introduced by the high-visibility scheme (see Fig. 4(a,d)). On the contrary, in the uniform-ZPD scheme and in the upgraded microscope, the largest phase shift is $\pm\pi/16$ (see Fig. 4(b,c,e,f)), which corresponds to $|\Delta t| = T/32$; this would lead to $\Delta \widetilde{H} \cong \Delta H \cdot 0.98$ (2% reduction).

An important parameter of the synthetic filter is its spectral width; this depends on the number $N$ of integrated samples, which corresponds to windowing the ideal interferogram by a rectangle with width $NT$ (see Fig. 11(a,c)). Eqs. (1-2) are still valid, taking into account that $G(f)$ would be the convolution between the FT of the whole interferogram and the FT of the window, that is the function $C(f)$ in Fig. 11(d). $C(f)$ hence represents the profile of the synthetic filter, the FWHM of which is of the order of $1/NT$, or equivalently, $f_0/N$, and its peak is $NT$. Note that since the convolved $G(0)$ also grows with N$T$, a large number of samples N would saturate the detector when acquiring $H_Q$ and $H_P$ (see Eq. (2)). This poses a trade-off between bandwidth and integrated amplitude for the choice of the optimal number of samples.

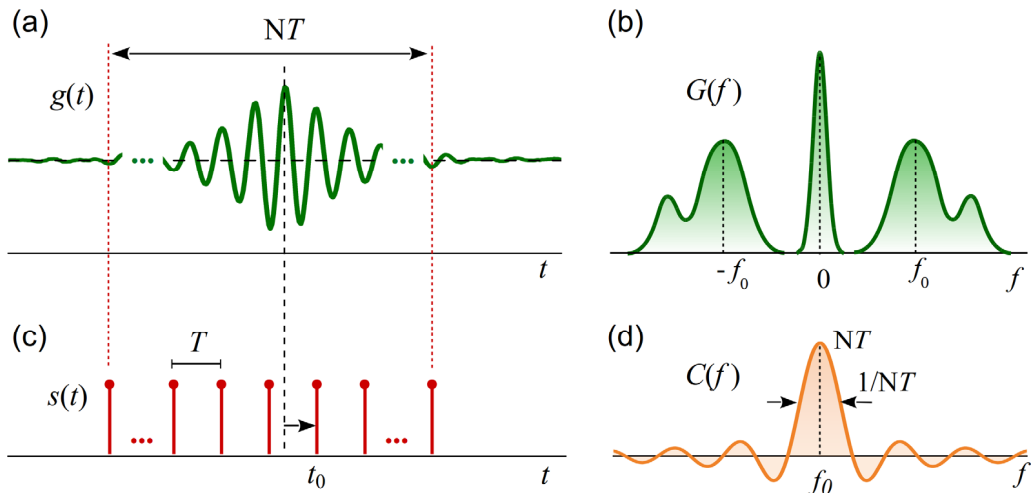

**Figure 11**: schematic representation of the sampling procedure; (a) *g(t)*: interferogram; (b) *G(f)*: Fourier transform of the interferogram; (c) *s(t)*: sampling function; T: sampling step; $t_0$: initial delay of the sampling, with respect to the interferogram time-zero. (d) *C(f)*: Shape of the synthetic spectral filter. Panels (b) and (d) are not on scale.

As shown, this method requires the collection of datapoints at two well defined sets of the delay. Since our uniform-ZPD HSM and our upgraded microscope enable controlling the delay with accuracy better than $\pm\pi/16$ across the whole detector plane (see Fig. 4(c,f)), both schemes are suited to apply the synthetic filter, hence providing band-passed images. In this case, instead of collecting the In-Phase and Quadrature datapoints, it is necessary to acquire two maps, also named In-Phase and Quadrature. In practice, this can be performed by continuously changing the interferometer delay and modulating the illumination/detection optical path with an on-off light shutter, so that the sample is illuminated only at the specific sampling delays. This has the advantage of reducing the light dose to the sample, which is important mainly with biological specimens. Each map is hence the result of the multiple exposure of the 2D images at the given delays (even multiples of T/2 for the In-Phase map, odd multiples of T/2 for the Quadrature). By subtracting the Quadrature from the In-Phase map, the band-passed image at $f_0$ is obtained.

Funding

BENF, AMdP and CM acknowledge financial support from the Brazilian funding agencies Capes (process CSF-PVE-S-88881.068168/2014-01), Fapemig and CNPq.

References

[1] S.W. Hell and J. Wichmann, Opt. Lett. **19**, 780 (1994).
[2] J. Huisken, J. Swoger, F. Del Bene, J. Wittbrodt, and E.H.K. Stelzer, Science **305**, 13 (2004).
[3] M.B. Sinclair, D.M. Haaland, J.A. Timlin, and H.D.T. Jones, Appl. Opt. **45**, 6283 (2006).
[4] A.J. Radosevich, M.B. Bouchard, S.A. Burgess, R. Stolper, B. Chen, and E.M.C. Hillman, Opt. Lett. 2008 **33**, 2164 (2008).
[5] F.R. Bertani, L. Ferrari, V. Mussi, E. Botti, A. Costanzo, and S. Selci, Sensors **13**, 14523 (2013).

[6] J.W. Cha, D. Tzeranis, J. Subramanian, I. V. Yannas, E. Nedivi, and P.T.C. So, Opt. Express **22**, 21368 (2014).

[7] P. Wang, C.G. Ebeling, J. Gerton, and R. Menon, Opt. Commun. **324**, 73 (2014).

[8] A. Orth, M.J. Tomaszewski, R.N. Ghosh, and E. Schonbrun, Optica **2**, 654 (2015).

[9] M.E. Gehm, M.S. Kim, C. Fernandez, and D.J. Brady, Opt. Express **16**, 11032 (2008).

[10] Z. Zhang, B. Hu, Q. Yin, T. Yu, and Z. Zhang, Mod. Phys. Lett. B **33**, 1 (2019).

[11] W. Jahr, B. Schmid, C. Schmied, F.O. Fahrbach, and J. Huisken, Nat. Commun. **6**, 1 (2015).

[12] A.D. Elliott, L. Gao, A. Ustione, N. Bedard, R. Kester, D.W. Piston, and T.S. Tkaczyk, J. Cell Sci. **125**, 4833 (2012).

[13] J.G. Dwight and T.S. Tkaczyk, Biomed. Opt. Express **8**, 1950 (2017).

[14] L. Gao, R.T. Kester, N. Hagen, and T.S. Tkaczyk, Opt. Express **18**, 14330 (2010).

[15] J. Wu, B. Xiong, X. Lin, J. He, J. Suo, and Q. Dai, Sci. Rep. **6**, 24624 (2016).

[16] M.F. Duarte, M.A. Davenport, D. Takhar, J.N. Laska, T. Sun, K.F. Kelly, and R.G. Baraniuk, IEEE Signal Process. Mag. **25**, 83 (2008).

[17] V. Studer, J. Bobin, M. Chahid, H.S. Mousavi, E. Candes, and M. Dahan, Proc. Natl. Acad. Sci. **109**, E1679 (2012).

[18] S.P. Davis, M.C. Abrams, and J.W. Brault, *Fourier Transform Spectrometry* (Academic Press, 2001).

[19] P.B. Fellgett, J. Opt. Soc. Am. **39**, 970 (1949).

[20] P. Jacquinot, Rep. Prog. Phys. **23**, 267 (1960).

[21] Z. Malik, D. Cabib, R.A. Buckwald, A. Talmi, Y. Garini, and S.G. Lipson, J. Microsc. **182**, 133 (1996).

[22] C. Ba, J.-M. Tsang, and J. Mertz, Opt. Lett. **43**, 2058 (2018).

[23] W. Müller, M. Kielhorn, M. Schmitt, J. Popp, and R. Heintzmann, Optica **3**, 452 (2016).

[24] D.N. Wadduwage, V.R. Singh, H. Choi, Z. Yaqoob, H. Heemskerk, P. Matsudaira, and P.T.C. So, Optica **4**, 546 (2017).

[25] D. Brida, C. Manzoni, and G. Cerullo, Opt. Lett. **37**, 3027 (2012).

[26] C.A. Manzoni, D. Brida, and G.N.F. Cerullo, US 9,182,284 (2015).

[27] A. Oriana, J. Réhault, F. Preda, D. Polli, and G. Cerullo, J. Opt. Soc. Am. A **33**, 1415 (2016).

[28] A. Perri, F. Preda, C. D'Andrea, E. Thyrhaug, G. Cerullo, D. Polli, and J. Hauer, Opt. Express **25**, A483 (2017).

[29] J. Réhault, M. Maiuri, D. Brida, C. Manzoni, J. Helbing, and G. Cerullo, Opt. Express **22**, 9063 (2014).

[30] J. Réhault, R. Borrego-Varillas, A. Oriana, C. Manzoni, C.P. Hauri, J. Helbing, and G. Cerullo, Opt. Express **25**, 4403 (2017).

[31] A. Perri, B.E. Nogueira de Faria, D.C.T. Ferreira, D. Comelli, G. Valentini, F. Preda, D. Polli, A.M. de Paula, G. Cerullo, and C. Manzoni, Opt. Express **27**, 15956 (2019).

[32] C. Mazoni et al., Patent application IT2018000008171, filing date 23 Aug. 2018.

[33] S. Krause and T. Vosch, Opt. Express **27**, 8208 (2019).

[34] M. Francon, S. Mallick, and others, *Polarization Interferometers* (Wiley, 1971).

[35] R.G. Baraniuk, IEEE Signal Process. Mag. **24**, 118 (2007).

[36] A.K. Jain and R.C. Dubes, *Algorithms for Clustering Data* (Prentice Hall, 1988).

[37] G. Accorsi, G. Verri, M. Bolognesi, N. Armaroli, C. Clementi, C. Miliani, and A. Romani, Chem. Commun. **23**, 3392 (2009).

[38] A. Cesaratto, C. D'Andrea, A. Nevin, G. Valentini, F. Tassone, R. Alberti, T. Frizzi, and D. Comelli, Anal. Methods **6**, 130 (2014).

[39] T. Jinno and M. Okuda, IEEE Trans. Image Process. **21**, 358 (2012).